\documentstyle[aps,preprint,prd]{revtex}

\def\bfeps{\mbox{\boldmath$\epsilon$}}

\def\bp{{\bf p}}
\def\bk{{\bf k}}
\def\H2{H^{**}}
\def \to {\rightarrow}

\begin{document}
\draft
\title{Spin Alignment of Heavy Meson Revisited}
\author{J.P. Ma}
\address{Institute of Theoretical Physics,\\
Academia Silica, \\
P.O.Box 2735, Beijing 100080, China\\
e-mail: majp@itp.ac.cn} \maketitle

\begin{abstract}
Using heavy quark effective theory a factorized form for inclusive
production rate of a heavy meson can be obtained, in which the
nonperturbative effect related to the heavy meson can be
characterized by matrix elements defined in the heavy quark
effective theory. Using this factorization, predictions for the
full spin density matrix of a spin-1 and spin-2 meson can be
obtained and they are characterized only by one coefficient
representing the nonperturbative effect. Predictions for spin-1
heavy meson are compared with experiment performed at $e^+e^-$
colliders in the energy range from $\sqrt{s}=10.5$GeV to
$\sqrt{s}=91$GeV, a complete agreement is found for $D^*$- and
$B^*$-meson. There are distinct differences
from the existing approach and they are discussed.
\end{abstract}
\par
\vskip50pt

Heavy quark effective theory(HQET) is a powerful tool to study
properties of heavy hadrons which contain one heavy quark
$Q$\cite{HQET,Review}. With HQET it allows such a study starting
directly from QCD. HQET is widely used in studied of decays of
heavy hadrons. In comparison, only in a few works it is used to
study productions of heavy hadrons. In 1994 Falk and Peskin used
HQET to predict spin alignment of a heavy hadron in its inclusive
production\cite{FP}. In this talk we will reexamine the subject
and restrict ourself to the case of spin-1 meson.
\par
A spin-1 heavy meson $H^*$ is a bound state of a heavy quark
$Q$ and a system of light degrees of freedom in QCD, like gluons and light quarks.
In the work\cite{FP} the total angular momentum $j$ of the light system is taken as
$1/2$. In the heavy quark limit, the orbital angular momentum of $Q$ can be neglected
and only the spin of $Q$ contributes to the total spin of $H^*$. Once the heavy quark
$Q$ is produced, it will combine the light system into $H^*$. Because parity is conserved
in QCD, the probabilities for the light system with positive- and negative helicity is the
same. Therefore, one can predict the probabilities for production of $H^*$ with a left-handed
heavy quark $Q$ as:
\begin{equation}
  P(\bar B^*(\lambda=-1)):P(\bar B^*(\lambda=0)):
  P(\bar B^*(\lambda=1)) =\frac{1}{2} : \frac{1}{4} :0,
\end{equation}
where $\lambda$ is the helicity of $H^*$. These results are easily derived, however,
three questions or comments can be made to them:
\par
(a). In general the spin information is contained in a spin density matrix,
probabilities are the diagonal part of the matrix. The question is
how about the non-diagonal part. It should be noted that this part
is also measured in experiment.
\par
(b). It is possible that the light system can have the total
angular momentum $j=3/2$ to form $H^*$. One may argue that the
production of such a system is suppressed. Is it possible to
derive full spin density matrix without the assumption of $j=1/2$?
\par
(c). How can we systematically add corrections to the  approximation which lead to the results in Eq.(1)?
\par
To make responses to these questions let us look at a inclusive production of $H^*$ in
detail. In its inclusive production a heavy meson is formed with a heavy
quark $Q$ and with other light degrees of freedom, the light degrees
can be a system of light quarks and gluons. Because its large mass
$m_Q$ the heavy quark is produced by
interactions at short distance. Therefore the production can be
studied with perturbative QCD. The heavy quark, once produced, will
combine light degrees of freedom to form a hadron, the formation
is a long-distance process, in which
momentum transfers are small, hence the formed
hadron will carry the most momentum of the heavy quark. The
above discussion implies the production rate can be
factorized, in which the perturbative part is for
the production of a heavy quark, while the nonperturbative
part is for the formation. For the nonperturbative
part an expansion in the inverse of $m_Q$ can systematically
be performed in the framework of HQET. This type
of the factorization was firstly used in parton fragmentation
into a heavy hadron\cite{Ma}.
\par
In this talk we will not discuss the factorization in detail, the details
can be found in the work\cite{Ma1}. We directly give our results and
make a comparison with experiment. It should be noted that the factorization
can be performed for any inclusive production of $H^*$. Because the most experiments
to measure spin alignment are performed at $e^+e^-$-colliders, we present the results
for the inclusive production at $e^+e^-$-colliders. We consider the process
\begin{equation}
e^+({\bf p})+e^{-}(-{\bf p})\to H^*({\bf k}) +X,
\end{equation}
where the three momenta are given in the brackets. In the
process we assume that the initial beams are unpolarized.
We denote the helicity of $H^*$ as $\lambda$ and
$\lambda=-1,0,1$. All information about the polarization
of $H^*$ is contained in a spin density matrix, which may
be unnormalized or normalized, we will call them
unnormalized or normalized spin density matrix, respectively.
The unnormalized spin density matrix can be defined as
\begin{equation}
R(\lambda, \lambda',{\bf p},{\bf k})
 = \sum_X \langle H^*(\lambda)X\vert {\cal T}\vert e^+e^-\rangle
         \cdot \langle H^*(\lambda')X\vert {\cal T}\vert e^+e^-\rangle^*,
\end{equation}
where the conservation of the total energy-momentum and the spin
average of the initial state is implied. ${\cal T}$ is the
transition operator. The cross-section with a given helicity
$\lambda$ is given by:
\begin{equation}
\sigma(\lambda) = \frac{1}{2s} \int \frac {d^3k}{(2\pi)^3}
R(\lambda, \lambda,{\bf p},{\bf k}).
\end{equation}
From Eq.(3) the normalized spin density matrix is defined by
\begin{equation}
\rho_{\lambda\lambda'}({\bf p},{\bf k}) =
 \frac {R(\lambda, \lambda',{\bf p},{\bf k})}
    {\sum_\lambda R(\lambda, \lambda,{\bf p},{\bf k})}.
\end{equation}
It should be noted that the normalized spin density matrix
is measured in experiment. It is straightforward to perform
the mentioned factorization for the unnormalized spin density matrix
in the rest frame of $H^*$, which is related to the moving frame
only by a Lorentz boost. In the rest frame we can define
a creation operator for $H^*$:
\begin{equation}
 \vert H^*(\lambda)\rangle =a^\dagger(\lambda) \vert 0\rangle=
  \bfeps(\lambda)\cdot{\bf a}^\dagger \vert 0 \rangle .
\end{equation}
where $\bfeps(\lambda)$ is the polarization vector. In the rest frame
the field $h_v$ of the heavy quark $Q$ in HQET has two non-zero components.
We denote them as:
\begin{equation}
h_v(x)=\left(\begin{array}{c} \psi(x) \\ 0 \end{array}\right).
\end{equation}
With these notations we define two operators:
\begin{equation}
O(H^*) = \frac{1}{6}{\rm Tr} \psi a_i^\dagger a_i \psi^\dagger, \ \
O_s(H^*) = \frac{i}{12} {\rm Tr} \sigma_i \psi a^\dagger_j
   a_k \psi^\dagger \varepsilon_{ijk},
\end{equation}
where $\varepsilon_{ijk}$ is the totally antisymmetric tensor and
$\sigma_i(i=1,2,3)$ is the Pauli matrix. The results for the unnormalized
spin density matrix read:
\begin{eqnarray}
R(\lambda, \lambda',{\bf p},{\bf k}) &=&\frac{1}{3} a({\bf p}, {\bf k})
 \langle 0 \vert O(H^*) \vert 0 \rangle \bfeps^*(\lambda)
 \cdot\bfeps(\lambda')
 \nonumber\\
  &+& \frac{i}{3} {\bf b}({\bf p},{\bf k})\cdot [\bfeps^*(\lambda)\times\bfeps(\lambda')]
\cdot \langle 0 \vert O_s(H^*) \vert 0 \rangle +{\cal O}(m_Q^{-2}).
\end{eqnarray}
The quantities $a({\bf p}, {\bf k})$ and ${\bf b}({\bf p},{\bf k})$ characterize
the spin density matrix of the heavy quark $Q$ produced in the inclusive process:
\begin{equation}
e^+({\bf p})+e^{-}(-{\bf p})\to Q({\bf k},{\bf s}) +X
\end{equation}
where ${\bf s}$ is the spin vector of $Q$ in its rest frame and
the rest frame is related to the moving frame only by a Lorentz
boost. The unnormalized spin density matrix $R_Q({\bf s},{\bf
p},{\bf k})$ of $Q$ can be defined by replacing $H^*(\lambda)$
with $Q({\bf k},{\bf s})$ in Eq.(3). This matrix can be calculated
with perturbative theory because of the heavy mass. The result in
general takes the form
\begin{equation}
R_Q({\bf s},{\bf p},{\bf k}) =a({\bf p},{\bf k}) +{\bf b}({\bf
p},{\bf k}) \cdot {\bf s}
\end{equation}
where $a({\bf p},{\bf k})$ and ${\bf b}({\bf p},{\bf k})$ are the
same in Eq.(9). The physical interpretation for Eq.(9) is the following:
The coefficients $a({\bf p},{\bf k})$ and ${\bf b}({\bf p},{\bf k})$
characterize the production of $Q$ and they can be calculated
with perturbative QCD,  while the two matrix elements defined
in HQET characterize the nonperturbative effects of the formation
of $H^*$ with the heavy quark $Q$.
With Eq.(9) we obtain:
\begin{equation}
\rho({\bf p},{\bf k}) =\frac{1}{3}\left(
\begin{array}{ccc}
  1+P_3, & -P_+, & 0 \\
   -P_-, & 1, & -P_+ \\
  0, &-P_-,  &1-P_3
\end{array}\right),
\end{equation}
with
\begin{equation}
P_3= \frac{b_3(\bp,\bk)}{a(\bp,\bk)}
            \cdot \frac{\langle 0 \vert O_s(H^*) \vert 0 \rangle}
             {\langle 0 \vert O(H^*) \vert 0 \rangle},\ \ \
P_\pm = \frac{b_1(\bp,\bk)\pm ib_2(\bp,\bk)}{\sqrt{2} a(\bp,\bk)}
           \cdot \frac{\langle 0 \vert O_s(H^*) \vert 0 \rangle}
             {\langle 0 \vert O(H^*) \vert 0 \rangle},\ \ \
\end{equation}
The indices of the matrix in Eq.(12) run from -1 to 1. Without knowing
the coefficients and the matrix elements we can already predict that
$\rho_{00}=1/3$ and $\rho_{1-1}=\rho_{-11}=0$.
\par
With these results we are in position to compare with experiment.
The experiments to measure the polarization of $B^*$ are performed
at LEP with $\sqrt s =M_Z$ by different experiments groups. To
measure the polarization the dominant decay $B^*\to \gamma B$ is used,
where the polarization of the photon is not observed. Because
the parity is conserved and the distribution of the angle
between the moving directions of $\gamma$ and of $B^*$ is measured,
one can only determine the matrix element $\rho_{00}$. If we denote
$\theta$ is the angle between the moving directions
of $B^*$ and of $\gamma$ in the $B^*$-rest frame and $\phi$
is the azimuthal angle of $\gamma$, then the angular distribution
is given by $W_{B^*\to B\gamma}(\theta,\phi)\propto \sum_{\lambda\lambda'}
\rho_{\lambda\lambda'}(\delta_{\lambda\lambda'}-Y_{1\lambda}(\theta,\phi)
Y^*_{1\lambda'}(\theta,\phi))$. Integrating over $\phi$ and using our result
$\rho_{00}=1/3$,
the distribution of $\theta$ is isotropic. In experiment one indeed finds that the
distribution is isotropic in $\theta$.
The experimental results at $\sqrt s =M_Z$ are\cite{DB,AB,OB}:
\begin{eqnarray}
 \rho_{00} &=& 0.32\pm 0.04\pm0.03,\ \ \ \ {\rm
 DELPHI} \nonumber \\
 \rho_{00}&=&0.33\pm0.06\pm0.05,\ \ \ \
 {\rm ALEPH}, \nonumber \\
 \rho_{00}&=&0.36\pm0.06\pm0.07,\ \ \ \
 {\rm OPAL}.
\end{eqnarray}
These results agree well with our prediction $\rho_{00}=1/3$.
\par
The polarization measurement for $D^*$-meson
has been done with different $\sqrt s$, in some
experiments the non-diagonal part of the spin density matrix has also been
measured by measuring azimuthal angular distribution in $D^*$ decay,
where the decay mode into two pseudo-scalars, i.e.,
$D^*\to D\pi$, is used. Denoting
$\theta$ as the angle between the moving directions
of $D^*$ and of $\pi$ in the $D^*$-rest frame and $\phi$
as the azimuthal angle of $\pi$, then the angular distribution of $\pi$
is given by $W_{D^*\to D\pi}(\theta,\phi)\propto \sum_{\lambda\lambda'}
\rho_{\lambda\lambda'}Y_{1\lambda}(\theta,\phi)
Y^*_{1\lambda'}(\theta,\phi)$. Integrating over $\phi$ and using our result
$\rho_{00}=1/3$,
the distribution of $\theta$ is again isotropic. The experimental results
are summarized in Table 1 and also partly summarized in \cite{THK}.
\par
From Table 1. we can see that the $\rho_{00}$ measured
by all experimental groups
is close to the prediction $\rho_{00}=\frac{1}{3}$,
the most precise result is obtained by CLEO, its
deviation from the prediction is $2\%$, the largest
deviation of the prediction is from the result made by OPAL
at $\sqrt s =90$GeV, it is $20\%$. In general, $\rho_{00}$
depends on the energy of $H^*$. Our results in Eq.(17) give
that $\rho_{00}$ is a constant in the heavy quark limit,
or the energy dependence is suppressed by $m_Q^{-2}$.
In experiment only a very weak energy dependence
is observed, e.g., in CLEO results\cite{CLEO}.
From our results $\rho_{1-1}$
is exactly zero in the heavy quark limit, the results from
TPC and from HRS  are in consistent with our result, a
non zero value is obtained by OPAL, which has a $3\sigma$
deviation from zero. These deviations may be explained
with effects of higher orders in $m_c^{-1}$, these
effects are expected to be substantial, because
$m_c$ is not so large. It is interesting to note only
results from OPAL at $\sqrt s =91$GeV have the largest
deviations from our predictions, while results from other
groups agree well with our predictions. At
$\sqrt s =10.5{\rm GeV\ or\ } 29$GeV, the effect of
the $Z$-boson exchange can be neglected, hence
the parity is conserved. We obtain $\rho_{10}=0$.
This prediction is also in agreement with the experimental result
made by TPC and by HRS.
\par\vskip20pt
\begin{center}
\centerline{{\bf Table 1}. Experimental Results for $D^*$}
\vspace{15pt}
\begin{tabular}{lll}
\hline\hline
  Collaboration\hspace{1cm} & $\sqrt s $\ in GeV\hspace{1.5cm} &  Results \\
  \hline
  CLEO\cite{CLEO} & 10.5  & $\rho_{00}=0.327\pm0.006$ \\
  HRS\cite{HRS} & 29    & $\rho_{00}=0.371\pm0.016$\\
  \ \           & \ \   & $\rho_{1-1}=0.04\pm0.03$ \\
  \ \           & \ \   & $\rho_{10}=0.00\pm0.01$ \\
  TPC\cite{TPC} & 29    & $\rho_{00}=0.301\pm0.042\pm0.007$\\
  \ \           &       & $\rho_{1-1}=0.01\pm0.03\pm0.00$ \\
  \ \           &       & $\rho_{10}=0.03\pm0.03\pm0.00$\\
  SLD\cite{SLD} & 91    & $\rho_{00}=0.34\pm0.08\pm0.13$\\
  OPAL\cite{OB} & 91    & $\rho_{00}=0.40\pm0.02\pm0.01$\\
  \ \           &  \ \  & $\rho_{1-1}=-0.039\pm0.014$\\
  \hline
\end{tabular}
\end{center}
\par\vskip15pt
Since our results are derived without knowing the total
angular momentum $j$ of the light degrees of freedom in the heavy
meson, the agreement of our results with experiment
can not be used to extract the information
about $j$ from the experimental data in Eq.(19) and in Table 1.,
although $\rho_{00}=1/3$ can also be obtained by taking $j=1/2$.
One way to extract $j$ may be to measure the difference between
$\rho_{11}-\rho_{-1-1}$, but it seems not
possible, because the polarization
of $H^*$ is measured through its parity-conserved decay and
the polarization of decay products is not observed in experiment.
In the heavy quark limit, the nondiagonal element $\rho_{1-1}$
and $\rho_{-11}$ are zero, while the other nondiagonal
matrix elements are nonzero if the parity is not conserved
and the initial state is unpolarized. At higher orders in $m_Q^{-1}$
this can be changed, e.g., $H^*$ can have tensor polarization.
\par
The factorization can also be done for inclusive productions of
a spin-2 meson. The results can be found in the work\cite{Ma1}.
Experimentally only the spin alignment of $D_2^*(2460)$ is
measured with large errors. But the experimental results seem
to be not in agreement with the predictions. The reason can be
the large effect from corrections at higher orders of $m_c^{1}$.
\par
To summarize: Using the approach of QCD factorization and employing
HQET, we obtain predictions of full spin density matrices of
spin-1- and spin-2 heavy meson. The leading order predictions for a spin-1 meson
agree well with experiment. Within the approach the three questions asked before
are answered. Although we have given in this talk detailed predictions
for inclusive production of a spin-1 heavy meson at
an $e^+e^-$ collider, our approach can be easily
generalized to other inclusive productions, testable predictions
can be made without a detailed calculation, for example, in inclusive
production of $B^*$ at an electron-hadron- or a hadron-hadron collider
we always have the prediction
$\rho_{00}=1/3$ and $\rho_{-11}=\rho_{1-1}=0$ in the heavy quark limit.

\end{document}